\documentclass{aa}  

\usepackage{graphicx}
\usepackage{txfonts}
\usepackage{hyperref}
\begin{document} 

\title{Phase-resolved {\it XMM-Newton} observations of the massive post-RLOF system HD\,149404\thanks{Based on data collected with {\it XMM-Newton}, an ESA Science Mission with instruments and contributions directly funded by the ESA Member States and the USA (NASA).}}
\author{G.\ Rauw\inst{1} \and S.\ Lizin\inst{1} \and S.\ Rosu\inst{2} \and E.\ Mossoux\inst{1}}
\institute{Space sciences, Technologies and Astrophysics Research (STAR) Institute, Universit\'e de Li\`ege, All\'ee du 6 Ao\^ut, 19c, B\^at B5c, 4000 Li\`ege, Belgium \and Department of Physics, KTH Royal Institute of Technology, The Oskar Klein Centre, AlbaNova, SE-106 91 Stockholm, Sweden\\ \email{g.rauw@uliege.be}}
\date{}

  \abstract
      {We investigated the X-ray emission of HD\,149404, a 9.81-day period O-star binary in a post-Roche lobe overflow evolutionary stage. X-ray emission of O-star binaries consists of the intrinsic emission of the individual O stars and a putative additional component arising from the wind-wind interaction.}{Phase-locked variations in the X-ray spectra can be used to probe the properties of the stellar winds of such systems.}{{\it XMM-Newton} observations of HD\,149404 collected at two conjunction phases and a quadrature phase were analysed. X-ray spectra were extracted and flux variations as a function of orbital phase were inferred. The flux ratios were analysed with models considering various origins for the X-ray emission.}{The highest and lowest X-ray fluxes are observed at conjunction phases respectively with the primary and secondary star in front. The flux variations are nearly grey with only marginal energy dependence. None of the models accounting for photoelectric absorption by homogeneous stellar winds perfectly reproduces the observed variations. Whilst the overall X-ray luminosity is consistent with a pure intrinsic emission, the best formal agreement with the observed variations is obtained with a model assuming pure wind-wind collision X-ray emission.}{The lack of significant energy-dependence of the opacity most likely hints at the presence of optically thick clumps in the winds of HD\,149404.}
\keywords{stars: early-type -- stars: individual (HD\,149404) -- binaries: close -- X-rays: stars}
\maketitle

\section{Introduction}
Over the past two decades, observational studies of the multiplicity of massive O-type stars has led to the conclusion that binarity and multiplicity are far more widespread among these stars than previously thought. As a result, binary interactions could play a predominant role in the evolution of massive stars \citep[e.g.][]{SF}. An interesting object in this context is HD\,149404. This system consists of an O7.5\,I(f) primary and an ON9.7\,I secondary on circular orbits with a period of 9.81\,days \citep{Rauw01}. \citet{Raucq} used spectral disentangling to reconstruct the spectra of the individual binary components and subsequently analysed these spectra with the CMFGEN model atmosphere code \citep{HM}. The authors found that the ON9.7\,I secondary star exhibits a nitrogen overabundance with [N/C] $\sim 100$\,[N/C]$_{\odot}$ that can only be explained if the system has gone through an episode of Case A Roche lobe overflow, confirming previous suggestions by \citet{Pen99} and \citet{Thaller}. The secondary star of HD\,149404 thus displays an evolutionary status intermediate between an O-type star and a WN Wolf-Rayet star.

\citet{Rauw19} analysed BRIght Target Explorer (BRITE) and Solar Mass Ejection Imager (SMEI) photometry of the binary, finding ellipsoidal variations with a peak-to-peak amplitude near 0.04\,mag. Combining the results of these light curves with the {\it Gaia}-DR2 parallax allowed us to constrain the orbital inclination of HD\,149404  in the range $23^{\circ}$ -- $31^{\circ}$ and the Roche lobe filling factor of the secondary star to $\geq 0.96$.

The optical spectrum of HD\,149404 displays strong line profile variability in the H$\alpha$ emission line and in the emission part of the He\,{\sc ii} $\lambda$\,4686 line \citep{Rauw01,Thaller,Naze}. These variations were interpreted as the signature of a wind-wind interaction occurring between the stars. Evidence of such an interaction was also provided by the BRITE photometry that suggests the presence of a bright spot at the surface of the secondary \citep{Rauw19}. Such a feature could result from the heating of the secondary photosphere by the wind interaction or from the crash of the primary wind onto the secondary's photosphere.

In some massive binaries, the shock-heated gas in the wind-wind collision zone produces significant X-ray emission that manifests itself through an overluminosity and orbital modulation of the observed X-ray emission \citep{SBP,PP10,RaNa16}. However, the current generation of X-ray satellites also revealed that strong X-ray emissions from the wind interaction zone are the exception, rather than the rule. Most massive binaries actually do not display any significant excess emission above the average canonical level \citep[e.g.][]{Naze09}.  

From a theoretical viewpoint, HD\,149404 is a promising target when it comes to using the X-ray emission from the stellar wind interactions to probe the properties of the winds of the two binary components. HD\,149404 is among the 12 X-ray brightest O + O binary systems detected during the ROSAT All-Sky Survey \citep[RASS;][]{Ber96} and it should be one of the most straightforward cases to be studied that way. Actually, many of the X-ray brighter O-star binaries in the RASS are higher multiplicity systems (e.g.\ $\delta$\,Ori, \citealt{Opl23}; $\tau$\,CMa, \citealt{Mai20}; HD\,150136, \citealt{San13}), which complicates the interpretation of their X-ray emission. Several other X-ray bright O-star systems have wide eccentric orbits (e.g. $\iota$\,Ori, \citealt{Mar00}; 9\,Sgr, \citealt{9Sgr}) and display X-ray emission that is notably modulated by the changing orbital separation. On the contrary, for a binary system with a circular orbit, such as HD\,149404, the situation should be simpler as putative orbital variations in the observed X-ray emission are expected to arise mostly from occultation effects and from the changing line-of-sight column density towards the shock region as the stars revolve around each other.

To study the X-ray properties of HD\,149404, we collected three {\it XMM-Newton} observations at key orbital phases to probe the properties of the stellar winds of this system and their interaction. In Sect.\,\ref{sect:obs} we present these new observations and their processing. The X-ray spectra are analysed in Sect.\,\ref{sect:spec}, and their variations with orbital phase are modelled under different assumptions in Sect.\,\ref{sect:model}. Our conclusions are presented in Sect.\,\ref{sect:conclusion}.

\begin{table*}
  \caption{{\it XMM-Newton} observations of HD\,149404.}
  \begin{center}
  \begin{tabular}{c c c c c c c}
    \hline
    Obs. & Obs.\ ID. & Duration & JD-2\,458\,000 & $\phi$ & Orbital config. & CR\\
    & & (ks) & & & & (ct\,s$^{-1}$)\\
    \hline
    1 & 0820510401 & 24.4 & 346.737 -- 347.089 & 0.209 -- 0.244 & Quadrature &$0.315 \pm 0.004$\\
    2 & 0820510501 & 26.0 & 354.222 -- 354.632 & 0.971 -- 0.013 & ON9.7I star in front & $0.233 \pm 0.003$\\
    3 & 0820510601 & 21.5 & 359.125 -- 359.437 & 0.471 -- 0.502 & O7.5I(f) star in front & $0.339 \pm 0.004$\\
    \hline
  \end{tabular}
  \end{center}
  \tablefoot{The duration indicated in the third column corresponds to the effective exposure time of the EPIC-pn camera (i.e.\ after removing the time interval affected by a flare in the second observation). The orbital phase range quoted in the fifth column was computed using the photometric ephemeris of \citet{Rauw19}: HJD$(\phi = 0) = 2\,457\,215.9943 + (9.81475 \pm 0.00084)\,E$. The count rate in the last column corresponds to the net (i.e.\ background-corrected) value recorded with the EPIC-pn camera within the spectral extraction region over its full energy band.\label{tab:obs}}
\end{table*}

\section{Observations and data processing \label{sect:obs}}
HD\,149404 was observed with the {\it XMM-Newton} satellite \citep{Jansen} on three occasions in August 2018 (see Table\,\ref{tab:obs}). The times of the observations were chosen to cover the two conjunction phases and one quadrature phase of the binary. {\it XMM-Newton} features three mirror modules that focus X-rays onto three European Photon Imaging Camera \citep[EPIC;][]{MOS,pn} instruments and two Reflection Grating Spectrometer \citep[RGS;][]{RGS} devices (RGS1 and RGS2). The EPIC instrument consist of two Metal Oxide Semi-conductor CCDs \citep[MOS;][]{MOS} and one p-n junction CCD \citep[pn;][]{pn}.

Our observations of HD\,149404 were taken with the EPIC cameras operated in full frame mode. Owing to the optical brightness of the source, we used the thick filters to prevent contamination of the X-ray signal by optical and UV photons. The first 10\%\ of the second observation was affected by a soft proton background flare and was thus discarded. We processed the data with the Science Analysis System (SAS) software version 18.0.0 using the current calibration files available in February 2020. Whilst there are several weak secondary sources in the EPIC field of view, HD\,149404 is by far the brightest object in the field and is well isolated from neighbouring X-ray sources.  

The EPIC spectra of the source were extracted from a circular source region  centred on the SIMBAD coordinates of our target and adopting a radius of 22\farcs5. The background was extracted from an annular region surrounding the source and with an outer radius of 30\,\arcsec. We also extracted background-corrected light curves of HD\,149404 over the 0.5\,keV -- 4.0\,keV band for each of the three EPIC instruments and for each of the three observations with temporal bin sizes of 100 and 1000\,s. We performed $\chi^2$ tests to assess the significance of short-term variability adopting different null hypotheses: a constant count rate, a linear trend with time, or a quadratic trend with time. In the majority of the light curves we found no significant variations. During our first observation, the EPIC-pn data showed significant variability consistent with a linear increase of few per cent  over the duration of the exposure. However, no significant variation was found in the MOS1 and MOS2 light curves recorded during this same observation. For the second observation, the 100\,s bin MOS1 and MOS2 light curves suggested marginally significant variability, which was not confirmed by the 1000\,s bin light curves of these instruments nor by the EPIC-pn light curves. Hence, we conclude that in most cases no significant intra-pointing variations were observed beyond the fluctuations consistent with Poisson noise. In addition to the EPIC spectra, we extracted first- and second-order grating spectra of HD\,149404 for both RGS instruments.

\section{Spectral fitting \label{sect:spec}}
\begin{figure}[htb]
\begin{center}
\resizebox{8cm}{!}{\includegraphics{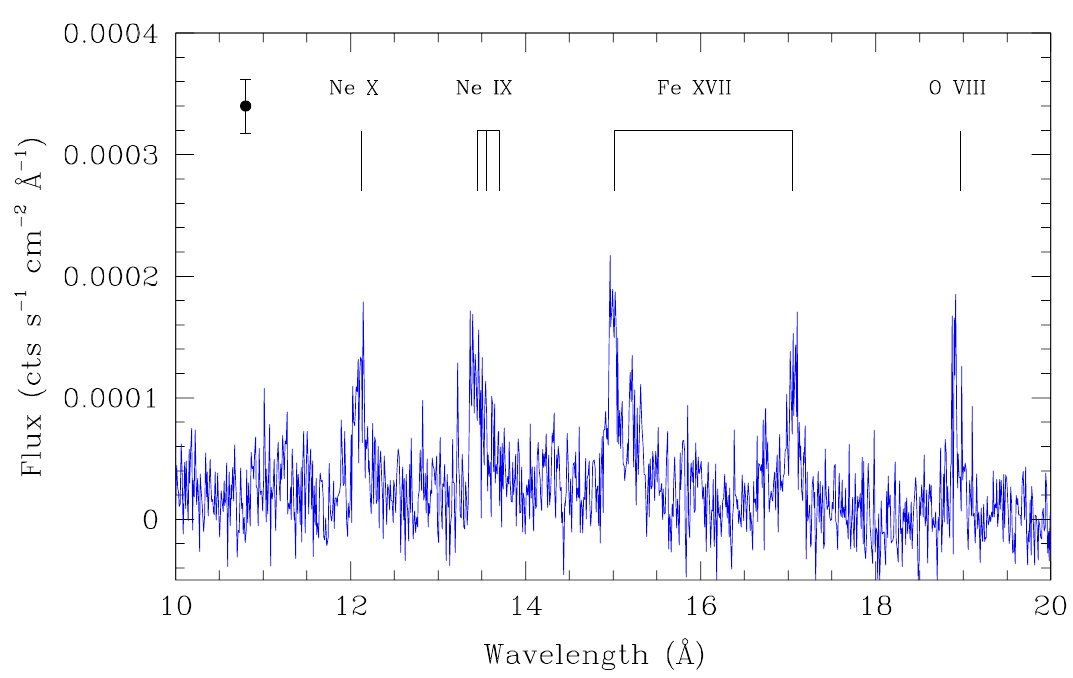}}
\end{center}
\begin{center}
\resizebox{8cm}{!}{\includegraphics{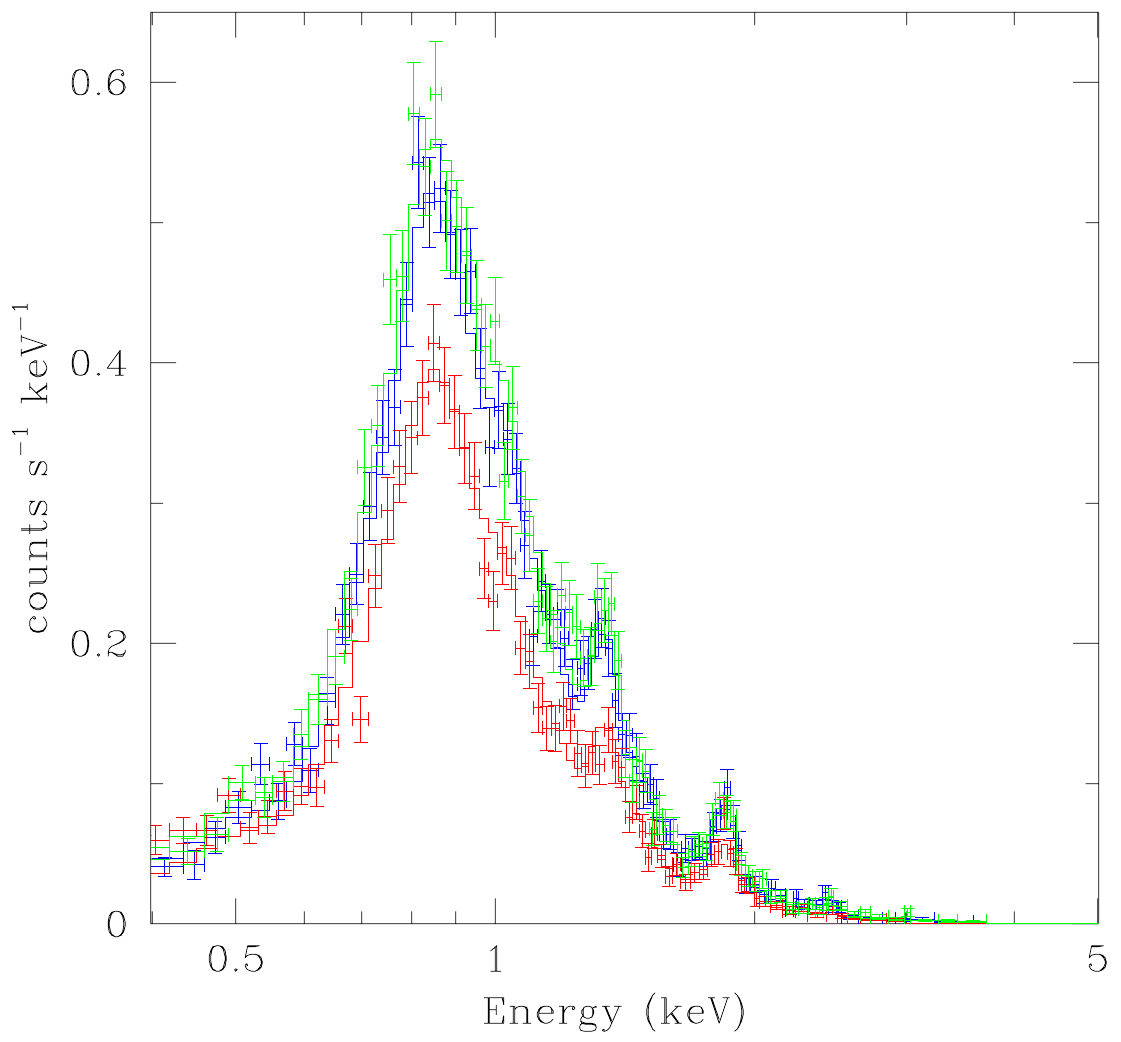}}
\end{center}
\caption{X-ray spectra of HD\,149404. \textit{Top panel}: Combined RGS (RGS1 + RGS2, first and second orders, all three observations) spectrum of HD\,149404. The labels indicate the lines that are apparent in this spectrum. The error bar in the top left corner indicates the mean error on the energy bins in the combined RGS spectrum between 10\,\AA\ and 20\,\AA. \textit{Bottom panel}: Comparison of the EPIC-pn spectra of HD\,149404 during the three observations. The first observation ($\phi = 0.227$) is shown in blue, the second ($\phi = 0.992$) in red, and the third ($\phi = 0.487$) in green. \label{RGScomb}}
\end{figure}

The X-ray spectrum of HD\,149404 displays emission lines that are characteristic of a thermal plasma (see Fig.\,\ref{RGScomb}). Several lines are clearly visible in the combined RGS data. These are Ne\,{\sc x} Ly\,$\alpha$ at 12.13\,\AA, the He-like {\it fir} triplet of Ne\,{\sc ix} at 13.70, 13.55, and 13.45\,\AA, the Fe\,{\sc xvii} lines at 15.01 and 17.05\,\AA, and the O\,{\sc viii} Ly\,$\alpha$ line at 18.97\,\AA. Whilst the quality of the RGS data is not sufficient to perform a detailed analysis of the line profiles for individual observations, the combined RGS spectra indicate that the {\it f} component of the Ne\,{\sc ix} triplet is suppressed, which is expected in the presence of a strong UV radiation field \citep{Por01} such as that  in the circumstellar environment of HD\,149404. No line is seen at longer wavelengths. This reflects a strong absorption by circumstellar and interstellar material, which can also be seen by the drop in the EPIC-pn spectra at energies below $\sim0.7$\,keV. Above photon energies of 4\,keV the EPIC spectra reveal no significant emission, suggesting that the X-ray emitting plasma is rather soft (see the bottom panel of Fig.\,\ref{RGScomb}).

To characterise the overall spectral energy distribution of HD\,149404's X-ray spectrum, we used version 12.9.0i of the {\tt xspec} code \citep{Arnaud}. We attempted to simultaneously fit the EPIC (i.e.\ MOS1, MOS2, and pn) and RGS (both orders of RGS1 and RGS2) spectra, using various combinations of absorbed optically thin thermal plasma models of the kind
\begin{equation}
  {\tt TBabs}*{\tt phabs}*\sum_{i=1}^n {\tt vapec}(kT_i).
  \label{fitexpression}
\end{equation}
In expression (\ref{fitexpression}) the interstellar absorption is accounted for by using the {\tt TBabs} model \citep{Wilms}. Based on the results of \citet{Dip94} and \citet{Rac09}, \citet{ism} quoted a total interstellar neutral hydrogen column density of $(3.72 \pm 0.94) \times 10^{21}$\,cm$^{-2}$. This value is in excellent agreement with the total neutral column density estimated from the reddening ($3.94 \times 10^{21}$\,cm$^{-2}$ for $E(B-V)=0.68$) using the $N($H\,{\sc i}$)/E(B-V)$ relation of \citet{Boh78}, and agrees reasonably well with the value ($4.16 \times 10^{21}$\,cm$^{-2}$) inferred from the $N($H\,{\sc i}$)/E(B-V)$ relation of \citet{ism}. We thus set the interstellar neutral hydrogen column density to $3.72 \times 10^{21}$\,cm$^{-2}$. Additional photoelectric absorption by the stellar winds is included by means of the {\tt phabs} model component, where the hydrogen column density is considered a free parameter of the model. The emission by the hot plasma is modelled by the sum of up to three collisionally ionised, optically thin {\tt vapec} thermal plasma components \citep{apec}. Such models are known to provide a good overall description of the X-ray spectral energy distribution of O-type stars \citep[e.g.][]{Naze09}. We explicitly requested the different {\tt vapec} plasma components to have identical chemical compositions.

Whilst the X-ray spectra of massive stars are frequently described by expressions such as (\ref{fitexpression}), this is certainly not a unique approach. For instance, one could also associate a particular circumstellar column density with each plasma component. However, there obviously exists some model degeneracy between the action of absorbing material and the intrinsic hardness of the various plasma components. Though we try to be as self-consistent as possible, we note that the primary goal of our {\tt xspec} fitting procedure is to obtain a good description of the overall X-ray spectral energy distribution. The specific parameters of the various model components should thus not be overinterpreted.  

\begin{table*}[htb]
  \caption{Spectral fits of the EPIC and RGS spectra using 2-T plasma models with a 50-50 mix of primary and secondary winds. \label{tab:fit2T}}
  \tiny
  \begin{center}
  \begin{tabular}{c c c c c c c c c}
    \hline
    Obs. & $N_{\rm H}$ & norm$_1$ & norm$_2$ & $\chi^2_{\nu}$ & d.o.f. & $f_X$ (soft) & $f_X$ (medium) & $f_X$ (hard) \\
    \cline{7-9}
    \vspace*{-2mm}\\
    & ($10^{22}$\,cm$^{-2}$) & (cm$^{-5}$) & (cm$^{-5}$) & & & \multicolumn{3}{c}{($10^{-14}$\,erg\,cm$^{-2}$\,s$^{-1}$)} \\
    \hline
    \vspace*{-2mm}\\
    1 & $0.48^{+0.04}_{-0.03}$ & $\left(4.9^{+1.4}_{-1.1}\right)\,10^{-3}$ & $\left(1.24^{+0.04}_{-0.04}\right)\,10^{-3}$ & 1.08 & 935 & $29.2^{+0.3}_{-0.5}$ & $31.5^{+0.5}_{-0.4}$ & $6.5^{+0.2}_{-0.1}$\\
    \vspace*{-2mm}\\
    2 & $0.41^{+0.03}_{-0.04}$ & $\left(2.6^{+0.7}_{-0.6}\right)\,10^{-3}$ & $\left(0.84^{+0.03}_{-0.03}\right)\,10^{-3}$ & 1.09 & 837 & $22.5^{+0.3}_{-0.4}$ & $22.2^{+0.3}_{-0.4}$ & $4.5^{+0.1}_{-0.1}$\\
    \vspace*{-2mm}\\
    3 & $0.51^{+0.04}_{-0.04}$ & $\left(6.7^{+1.8}_{-1.4}\right)\,10^{-3}$ & $\left(1.29^{+0.04}_{-0.04}\right)\,10^{-3}$ & 1.19 & 841 & $31.3^{+0.3}_{-0.5}$ & $33.4^{+0.5}_{-0.4}$ & $6.8^{+0.1}_{-0.1}$\\
    \vspace*{-2mm}\\
    \hline
  \end{tabular}
  \end{center}
  \tablefoot{The model was defined as Eq.\,(\ref{fitexpression}) with $n = 2$. The interstellar column was set to $3.72 \times 10^{21}$\,cm$^{-2}$ \citep{ism}, whereas the plasma temperatures were fixed to 0.20 and 0.65\,keV. The abundances of C, N, and O were set respectively to 0.22, 6.26, and 0.83 times solar. All other abundances were fixed to solar following \citet{Asplund}. The plasma normalisation parameters are equal to $\frac{10^{-14}}{4\,\pi\,d^2}\,\int n_e\,n_{\rm H}\,dV$ in units cm$^{-5}$ with $d$ the distance to the source and $\int n_e\,n_{\rm H}\,dV$ the emission measure of the plasma. The last three columns yield the observed fluxes in the soft (0.5\,keV -- 1.0\,keV), medium (1.0\,keV -- 2.0\,keV), and hard (2.0\,keV -- 4.0\,keV) energy bands.}
\end{table*}

\begin{table*}[htb]
  \caption{Spectral fits of the EPIC and RGS spectra using 3-T plasma models with a 50-50 mix of primary and secondary winds. \label{tab:fit3T}}
  \tiny
  \begin{center}
  \begin{tabular}{c c c c c c c c c c}
    \hline
    Obs. & $N_{\rm H}$ & norm$_1$ & norm$_2$ & norm$_3$ & $\chi^2_{\nu}$ & d.o.f. & $f_X$ (soft) & $f_X$ (medium) & $f_X$ (hard) \\
    \cline{8-10}
    \vspace*{-2mm}\\
    & ($10^{22}$\,cm$^{-2}$) & (cm$^{-5}$) & (cm$^{-5}$) & (cm$^{-5}$) & & & \multicolumn{3}{c}{($10^{-14}$\,erg\,cm$^{-2}$\,s$^{-1}$)} \\
    \hline
    \vspace*{-2mm}\\
    1 & $0.47^{+0.03}_{-0.04}$ & $\left(5.7^{+2.0}_{-1.6}\right)\,10^{-3}$ & $\left(1.7^{+0.2}_{-0.3}\right)\,10^{-3}$ & $\left(7.6^{+0.5}_{-0.5}\right)\,10^{-4}$ & 1.05 & 934 & $28.5^{+0.4}_{-0.4}$ & $32.1^{+0.4}_{-0.5}$ & $6.3^{+0.2}_{-0.2}$\\
    \vspace*{-2mm}\\
    2 & $0.41^{+0.04}_{-0.04}$ & $\left(4.3^{+1.4}_{-1.1}\right)\,10^{-3}$ & $\left(1.0^{+0.2}_{-0.2}\right)\,10^{-3}$ & $\left(5.5^{+0.4}_{-0.4}\right)\,10^{-4}$ & 1.05 & 836 & $22.2^{+0.3}_{-0.4}$ & $22.5^{+0.4}_{-0.3}$ & $4.5^{+0.2}_{-0.1}$\\
    \vspace*{-2mm}\\
    3 & $0.49^{+0.04}_{-0.03}$ & $\left(8.8^{+3.0}_{-2.3}\right)\,10^{-3}$ & $\left(1.8^{+0.3}_{-0.3}\right)\,10^{-3}$ & $\left(8.0^{+0.6}_{-0.6}\right)\,10^{-4}$ & 1.15 & 840 & $30.5^{+0.3}_{-0.5}$ & $34.1^{+0.4}_{-0.5}$ & $6.7^{+0.2}_{-0.2}$\\
    \vspace*{-2mm}\\
    \hline
  \end{tabular}
  \end{center}
  \tablefoot{Same as Table\,\ref{tab:fit2T}, but for the model defined as Eq.\,(\ref{fitexpression}) with $n = 3$. The plasma temperatures were fixed to 0.14, 0.35, and 0.73\,keV.}
\end{table*}

\begin{figure}[htb]
\begin{center}
\resizebox{8cm}{!}{\includegraphics{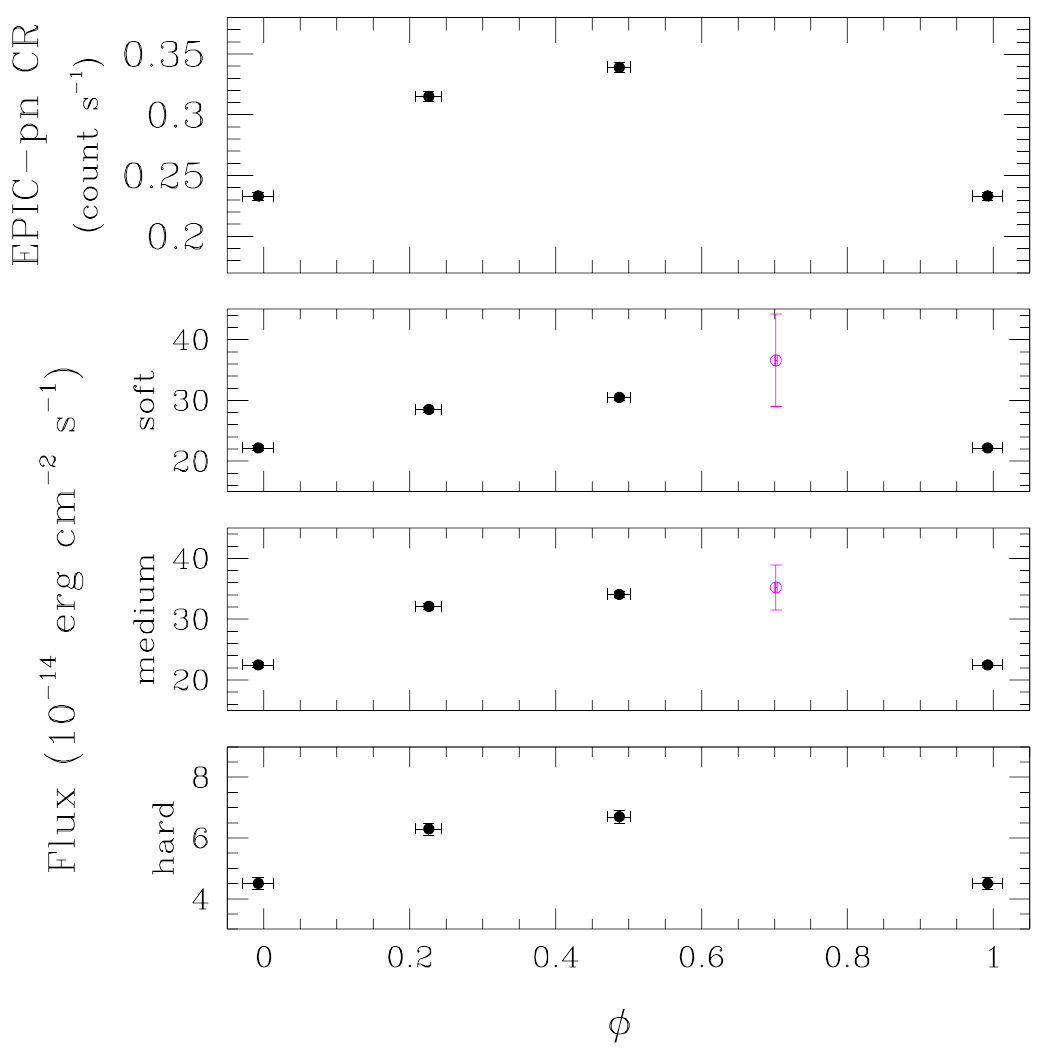}}
\end{center}
\caption{Variations of the X-ray emission of HD\,149404 with orbital phase. From top to bottom, the panels illustrate the full EPIC-pn count rates for the different observations and the observed fluxes evaluated over three different energy bands: soft (0.5\,keV -- 1.0\,keV), medium (1.0\,keV -- 2.0\,keV), and hard (2.0\,keV -- 4.0\,keV). The magenta symbols in the second and third panels correspond to the fluxes measured on an archival ROSAT-PSPC spectrum \citep{Rauw01}. \label{variabilityEPIC}}
\end{figure}

Our very first fitting trials indicated that a single plasma component is not sufficient to achieve a decent quality fit of the spectra. Good fitting qualities were instead achieved for models with two or three components. We set the chemical abundances to solar according to \citet{Asplund}, except for the C, N, and O abundances. Tests with these three abundances allowed to vary indicated that the spectral fits are essentially insensitive to the C abundance, whereas they systematically yielded a significant N overabundance and an O abundance consistent with solar. These results most likely indicate that the X-ray emitting plasma contains some gas from the secondary star. We thus tested several assumptions for the CNO abundances of the X-ray emitting plasma: solar composition, a pure primary wind composition, a 50-50 mix of primary and secondary wind compositions, and a pure secondary wind composition, the last three with abundances as inferred by \citet{Raucq}.

The quality of the fits significantly improved when allowing for non-solar composition. The best fit was achieved for an equal proportion mix of primary and secondary abundances ($\chi_\nu$ between 1.05 and 1.15; see Tables\,\ref{tab:fit2T} and \ref{tab:fit3T}). However, the fit quality significantly degraded ($\Delta \chi_{\nu}^2 \sim +0.20$) for a 100\% secondary composition of the emitting plasma. The reason for this is the low O abundance of the secondary wind inferred by \citet{Raucq}, which seems inconsistent with the EPIC spectra and with the presence of the O\,{\sc viii} Ly\,$\alpha$ line in the RGS spectra. A less extreme degradation of the fit quality ($\Delta \chi_{\nu}^2 \sim +0.05$) occurs if we assume a pure primary wind composition. These tests hence suggest that the X-ray emitting plasma in HD\,149404 contains a mix of material from both winds. Finally, we also tested models where the proportion of primary and secondary abundances was left as a free parameter. Those models yielded a secondary abundance contribution of $(46 \pm 12)$\%. The quality of these  fits was essentially the same ($|\Delta \chi^2_\nu| \leq 0.01$) as for the 50-50 mix models. In the remainder of this paper we thus focus on the results obtained for the equal proportion mix. 
    
For a two-temperature (2-T) plasma consisting of a 50-50 mix, we fitted the spectra of the three observations independently keeping the wind column density and the plasma temperatures as free parameters. The temperatures of the two plasma components were found to be very similar in all three observations: $kT_1 = 0.20 \pm 0.04$\,keV and $kT_2 = 0.65 \pm 0.06$\,keV. We thus repeated the fits, fixing the two plasma temperatures, but allowing their emission measures to vary from one observation to another (see Table\,\ref{tab:fit2T}).

We repeated the same procedure for a three-temperature (3-T) plasma model with chemical abundances consisting of a 50-50 mix. Again, the temperatures of the three plasma components were found to be very similar in all three observations: $kT_1 = 0.14 \pm 0.04$\,keV, $kT_2 = 0.35 \pm 0.07$\,keV, and $kT_3 = 0.73 \pm 0.02$\,keV. The results obtained with these temperatures fixed but allowing the emission measures and the wind column density to vary are quoted in Table\,\ref{tab:fit3T}. An F-test to assess whether  including a third {\tt vapec} component significantly improves the quality of the fit yields false alarm probabilities of $10^{-6}$, $10^{-8}$, and $6 \times 10^{-7}$ for the first, second, and third observations, respectively. Hence, we conclude that the third plasma component indeed provides an improvement in the fit quality.

This soft X-ray emission might be surprising at first sight for a putative interacting wind system, but it is not unusual for short period O-star binaries. For instance, {\it XMM-Newton} spectra of the O7.5\,III-II(f) + O7.5\,III-II(f) system HD\,152248 ($P_{\rm orb} = 5.8$\,d) were fitted by \citet{San04} with a dominant plasma component with $kT \sim 0.2$\,keV and a hottest plasma component with $kT \sim 0.8$\,keV. Likewise, for the {\it ASCA}-SIS spectra of the highly eccentric O9\,III + B1\,III system $\iota$\,Ori ($P_{\rm orb} = 29.1$\,d, $e = 0.76$), \citet{Pit00} inferred best-fit plasma temperatures of 0.15\,keV and 0.61\,keV.  

Over the 0.5 -- 4.0\,keV energy domain, the fluxes vary between $4.9 \times 10^{-13}$\,erg\,cm$^{-2}$\,s$^{-1}$ and $7.2 \times 10^{-13}$\,erg\,cm$^{-2}$\,s$^{-1}$. When correcting for the absorption by the interstellar material only, the corresponding fluxes are $1.09 \times 10^{-12}$\,erg\,cm$^{-2}$\,s$^{-1}$ and $1.53 \times 10^{-12}$\,erg\,cm$^{-2}$\,s$^{-1}$. Comparing these numbers to the bolometric fluxes of the system \citep{Raucq,Rauw19}, we obtain a distance-independent estimate of $\log{\frac{f_X}{f_{\rm bol}}} = -7.13 \pm 0.03$ at minimum and $-6.97 \pm 0.03$ at maximum. This result is very similar to the canonical value of O-type stars \citep{Naze09}, indicating that HD\,149404 is one of those O-star binaries that do not exhibit any strong X-ray overluminosity. If anything, the system would appear to be underluminous. We come back to this point in Sect.\,\ref{EWS}.

Together with the overall softness of the X-ray emission, these results suggest that HD\,149404 does not exhibit any strong X-ray excess emission arising from the shocked winds. This situation probably stems from the fact that the stars that make up this binary system are large compared to their orbital separation. As a result, the stellar winds do not have enough room to accelerate to high velocities before they interact. Therefore, the material in the wind interaction zone may be too cool to produce any strong and hard X-ray emission (see Sect.\,\ref{EWS}).

The variations in the X-ray spectrum and fluxes of HD\,149404 as a function of orbital phase are illustrated in Figs.\,\ref{RGScomb} and \ref{variabilityEPIC}. In Fig.\,\ref{variabilityEPIC} we also include the results of an archival ROSAT-PSPC observation, taken at phase $\phi = 0.70$, which was previously discussed in \citet{Rauw01}. This figure shows that the X-ray fluxes display rather limited variability, except for the drop around phase 0.0, which corresponds to the conjunction with the ON9.7\,I secondary being in front. At this phase, the line of sight towards the wind interaction zone mostly samples the wind of the ON9.7\,I secondary, whilst it probes the wind of the O7.5\,I(f) primary at the other conjunction phase 0.5. 

\begin{figure}[h]
\begin{center}
\resizebox{8cm}{!}{\includegraphics{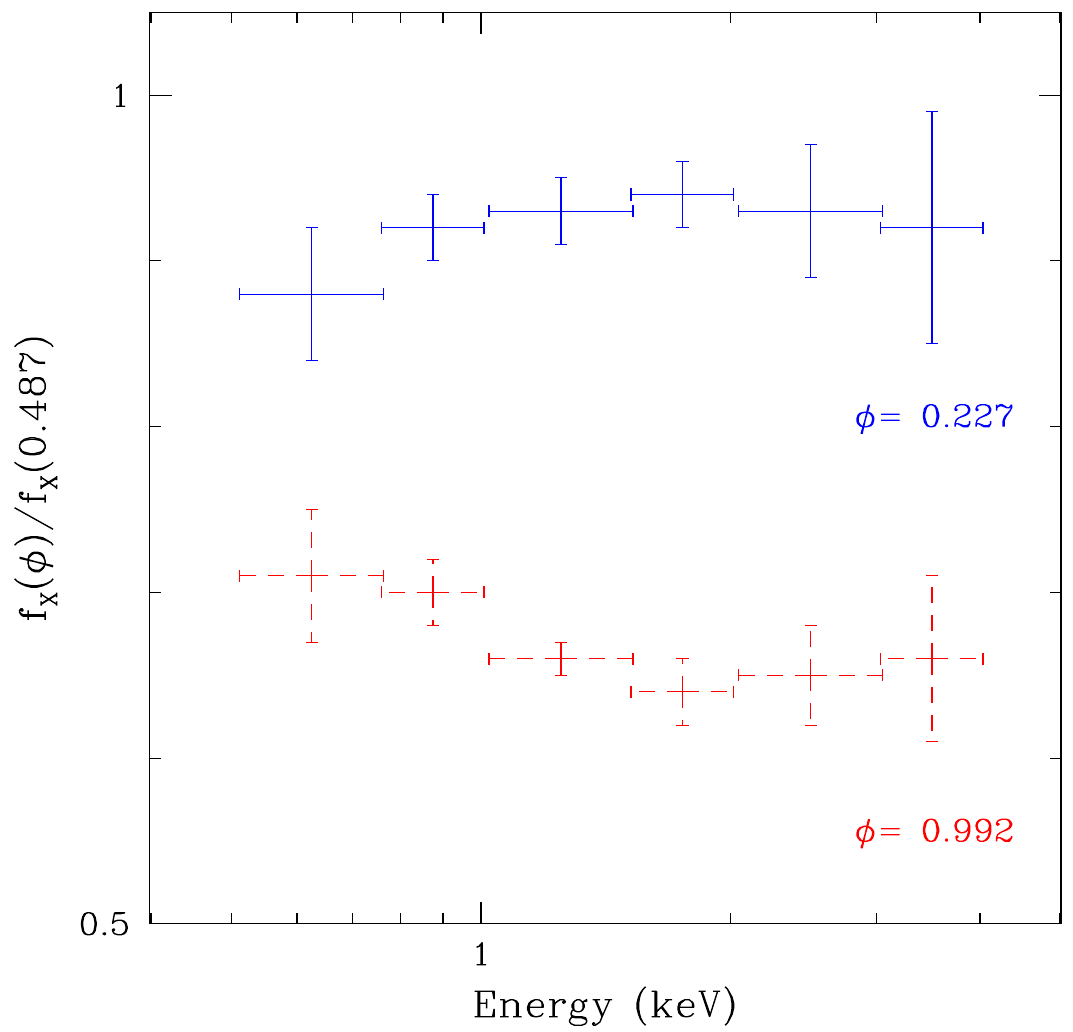}}
\end{center}  
\caption{Ratio of the observed fluxes evaluated over narrow energy bands. The blue (resp. red) symbols indicate the ratios of the fluxes measured at phase 0.227 (resp.\ 0.992) to those determined at phase 0.487, the phase at which the emission is strongest. \label{fig:fx}}
\end{figure}

It is worth noting that the time of minimum flux, around phase 0.0, corresponds to the lowest circumstellar column density in our {\tt xspec} fitting. This behaviour is also observed for the 2-T and 3-T plasma models. We note that the model parameters should not be overinterpreted, and that the formal errors on $N_{\rm H}$ do not allow us to rule out the possibility of a constant column density. However, our results suggest that the observed flux variation is not due to phase-modulated absorption column density. Instead, the fits in Tables \ref{tab:fit2T} and \ref{tab:fit3T} indicate that the lower flux at phase 0.0 stems from a global reduction of the emission measure of the emitting plasma at that phase (see also Fig.\,\ref{fig:fx}). This suggests that occultation of the emitting region plays an important role. 

\section{The variations in the X-ray emission \label{sect:model}}
Figure\,\ref{fig:fx} illustrates the ratio of the observed fluxes in various energy bands at the orbital phases sampled by our {\it XMM-Newton} observations. A remarkable property of these variations is that they are essentially grey (i.e.\ nearly independent of photon energy). In this section, we attempt using the observed variations in the X-ray spectrum between different orbital phases to probe the properties of the winds of HD\,149404. For this purpose, we consider various scenarios: either the observed X-ray emission is dominated by the wind interaction zone, or it is mostly intrinsic to the stellar winds with only a moderate contribution from the wind interaction zone.   
\subsection{X-ray opacity of the stellar winds \label{opacX}}
To estimate the X-ray opacities $\kappa_E$ of the stellar winds of HD\,149404's components, we used the CMFGEN non-LTE model atmosphere code \citep{HM} together with the stellar parameters and chemical compositions inferred by \citet{Raucq} and recalled in Table\,\ref{tab:paramCMFGEN}. Overall, the computed opacities of the two winds are rather similar, except for some differences in the region around 0.5\,keV, which are mainly due to the enhanced nitrogen and depleted carbon abundances of the secondary star. Over the energy domain between 0.5 and 4.0\,keV, which is of interest for our present study, the opacities vary relatively little with position inside the wind. It is only for the innermost positions in the wind that the opacity drops significantly owing to the higher ionisation stage of the wind material in these regions. We thus consider that the opacity at a distance from the stellar centres $r = a/2$, where $a$ is the semi-major axis of the orbit, is representative of the wind opacity over those regions that are most relevant to our present purpose. 

\begin{table}
  \caption{Stellar and wind parameters adopted in the computation of the X-ray opacities. \label{tab:paramCMFGEN}}
  \begin{center}
  \begin{tabular}{c c c }
    \hline
    & Primary & Secondary \\
    \hline
    $T_{\rm eff}$ (K) & 34\,000 & 28\,000 \\
    $R$ ($R_{\odot}$) & 15.7 & 21.8 \\
    $v_{\infty}$ (km\,s$^{-1}$) & 2450 & 2450 \\
    $\dot{M}$ ($M_{\odot}$\,yr$^{-1}$) & $9.2 \times 10^{-7}$ & $3.3 \times 10^{-7}$ \\
    $\beta$ & 1.03 & 1.08 \\
    $\epsilon_{\rm C}/\epsilon_{\rm C, \odot}$ & 0.38 & 0.07 \\
    $\epsilon_{\rm N}/\epsilon_{\rm N, \odot}$ & 1.95 & 10.6 \\
    $\epsilon_{\rm O}/\epsilon_{\rm O, \odot}$ & 1.50 & 0.16  \\
    \hline
  \end{tabular}
  \end{center}
  \tablefoot{The parameters are from \citet{Raucq}, except for the radii taken from the combined photometric and astrometric analysis of \citet{Rauw19}. The mass-loss rates were computed using the formalism of \citet{Muijres}.}   
\end{table}

\begin{figure}[h]
\begin{center}
\resizebox{8cm}{!}{\includegraphics{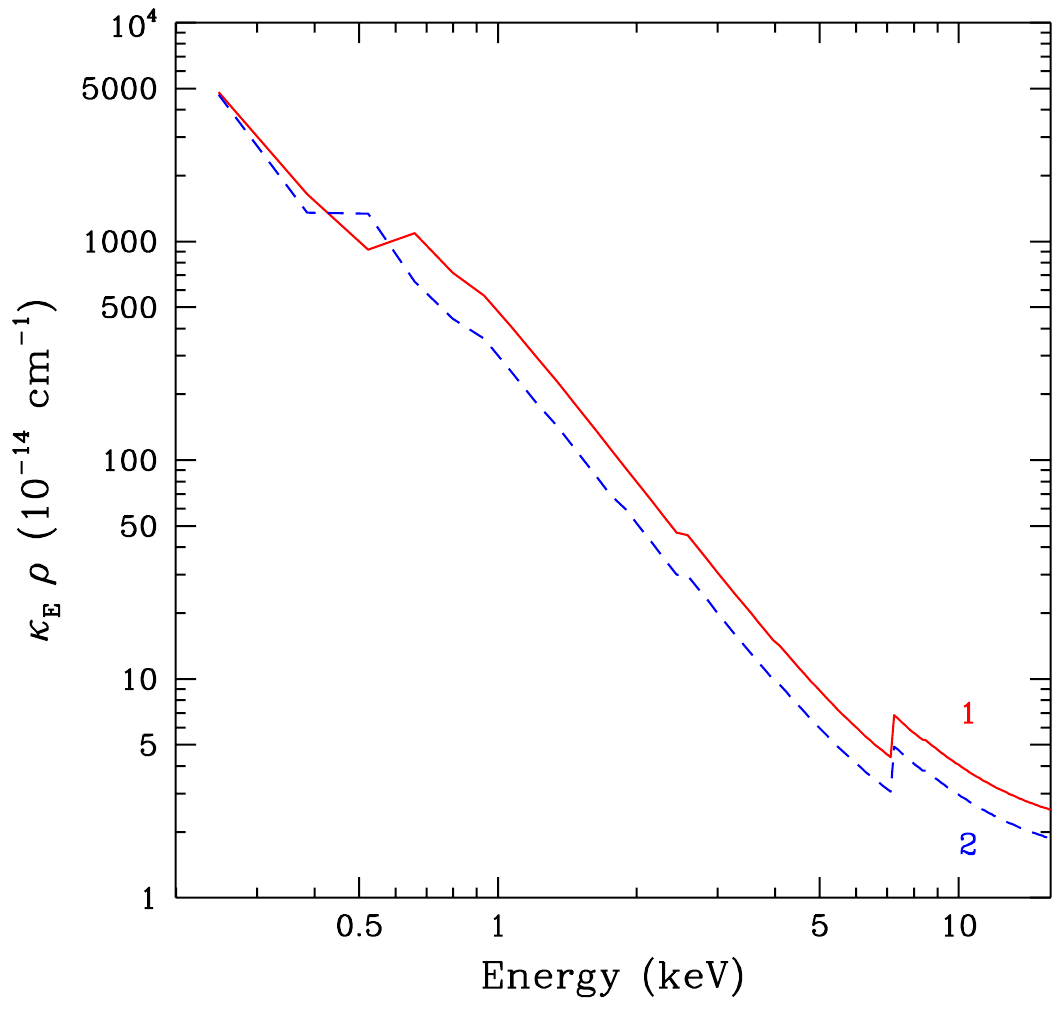}}
\end{center}  
\caption{Product of wind X-ray opacity $\kappa_E$ times wind density $\rho$ as a function of energy. The opacity and wind density are computed from the CMFGEN models of \citet{Raucq} at a distance $r = a/2$ from the stellar centre. The continuous red line (labelled 1) shows the results for the primary star, whereas the dashed blue line (labelled 2) corresponds to the secondary star of HD\,149404. \label{fig:kappa}}
\end{figure}

The optical depth along a given line of sight is obtained by integrating the product of $\kappa_E$ by the wind density $\rho$
\begin{equation}
\tau_E = \int \kappa_E\,\rho\,ds,
\end{equation}
where the integration starts from the position of the considered point in the wind interaction zone and proceeds along the line of sight towards the observer.
To compare the effects of the two winds, Fig.\,\ref{fig:kappa} shows the product of $\kappa_E$ by $\rho$ evaluated at a position $r = a/2$ in the wind. As is apparent, the CMFGEN models predict the primary wind to be slightly more opaque than the secondary wind at most energies. This would be in line with the marginally higher column density in the {\tt xspec} models around phase 0.5 compared to phase 0.0. However, our observational results (Figs.\,\ref{variabilityEPIC} and \ref{fig:fx}) indicate that the observed X-ray fluxes are lower at the conjunction with the secondary star being in front ($\phi = 0.0$), thus suggesting once more that the optical depth variations are not the main cause of the reduced observed flux at phase $0.0$. 

Whilst the values of $\kappa_E\,\rho$ of both stars are remarkably similar (Fig.\,\ref{fig:kappa}), they vary very strongly with energy: the values at 0.5 and 5.0\,keV differ by about a factor of 100. This provides another argument suggesting that the nearly grey attenuation of the observed fluxes seen in Fig.\,\ref{fig:fx} cannot stem from the sole effect of changing wind absorption, at least not if the emission at all energies arises from the same geometrical region. A possibility to produce a grey attenuation would be through a phase-dependent occultation effect of the emitting region by the stellar bodies. Adopting our best estimates of the stellar radii (Table\,\ref{tab:paramCMFGEN}), the surface area of the secondary star is about twice as large as that of the primary. If the X-ray emitting region is located very close to the secondary's surface facing the primary, we would expect a smooth phase-dependent modulation of the observable emission with the minimum flux occurring at phase $0.0$. Because of the low orbital inclination, the more extended the emission region, the smaller the expected variations in the X-ray flux between different orbital phases.   

\subsection{X-ray emission from the wind interaction zone \label{CWB}}
\begin{figure}[h]
\begin{center}
\resizebox{8cm}{!}{\includegraphics{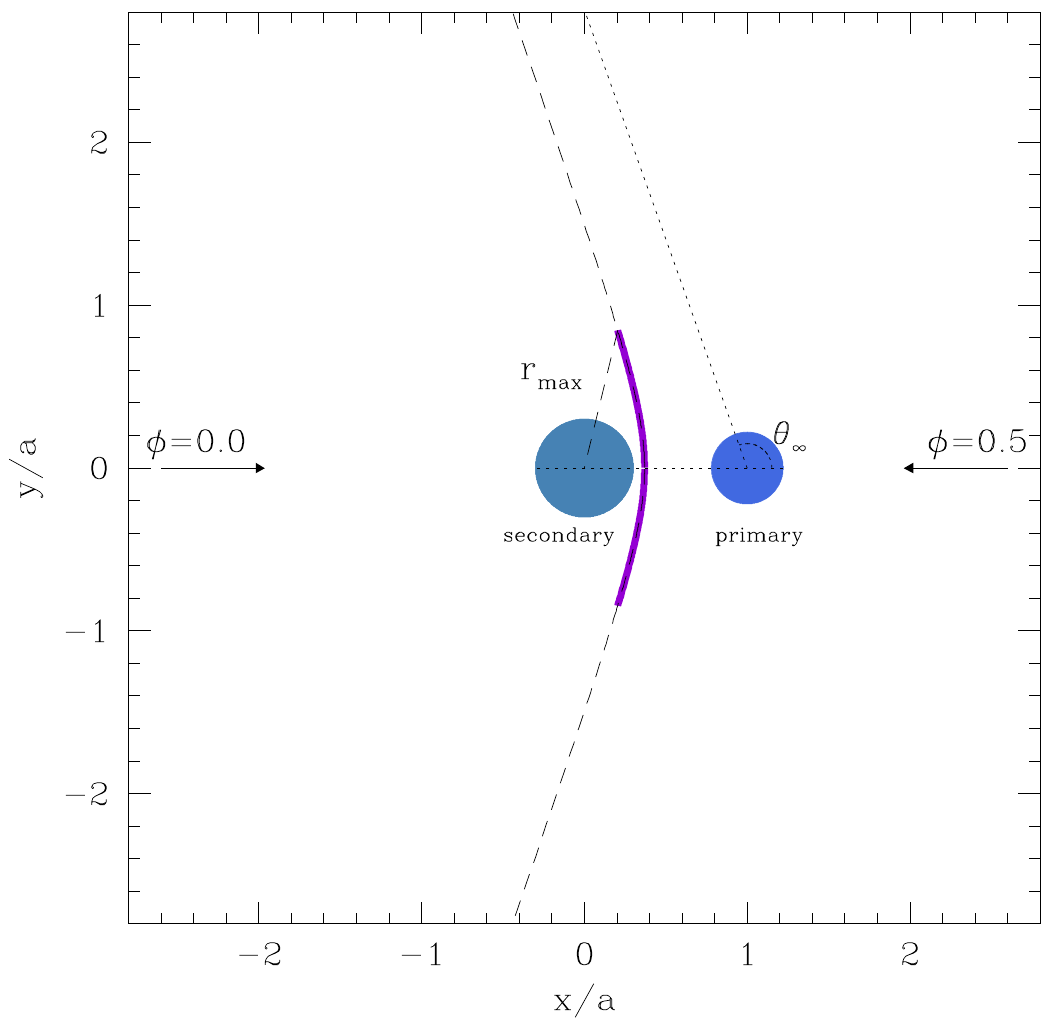}}
\end{center}  
\caption{Schematic illustration of the wind interaction zone in the orbital plane of HD\,149404 modelled with the formalism of \citet{Canto}. \label{fig:schema}}
\end{figure}

Although we  show in Sect.\,\ref{opacX} that the X-ray luminosity of HD\,149404 is rather modest and that the spectrum is soft, we start by considering as a first hypothesis that the observed X-ray emission of HD\,149404 is dominated by the wind interaction zone. For this part of the analysis, we followed the same approach as previously done by \citet{Rauw14} in the study of HDE\,228766. We used the formalism of \citet{Canto} to compute the shape of the axisymmetric wind interaction zone (see Fig.\,\ref{fig:schema}) and discretise the three-dimensional interaction region into a total of $200 \times 360$ cells along the $\theta$ and $\varphi$ spherical coordinates.

As a first step we computed the density and plasma temperature in the wind interaction zone and the associated X-ray emission by the shock-heated plasma in the same way as done in \citet{Rauw16}. This approach assumes that the plasma in the wind interaction zone cools adiabatically. We then computed the optical depth along the line of sight towards a given cell of material in the wind interaction zone (see Appendix\,\ref{Appendix}).

Our calculations further accounted for occultations of the plasma cells by either of the two stars, as well as the lines of sight that cross both winds. For the plasma emission, we used emissivities that are representative of prominent spectral lines that are present over the range of temperatures found in the fits of the {\it XMM-Newton} spectra (see Table\,\ref{tab:fit3T}). For instance, the emissivity $Q_E(T)$ in the 0.5 - 0.75\,keV energy band was taken to be that of the strong O\,{\sc viii} Ly\,$\alpha$ line at 0.65\,keV.

The relevant parameters in this model are the mass-loss rates $\dot{M}_1$ and $\dot{M_2}$ of the stars with the stronger and weaker wind respectively, the associated wind velocities $v_{\infty,1}$ and $v_{\infty,2}$, as well as the orbital inclination $i$. Taking advantage of the constraints on $i$ established from the BRITE and SMEI light curves \citep{Rauw19}, we restricted our investigation to inclinations in the range  $20^{\circ}$ -- $40^{\circ}$. The geometrical shape and the location of the shock region are defined by the wind-momentum ratio $\eta = \frac{\dot{M}_1\,v_1}{\dot{M}_2\,v_2}$ \citep{Canto}. The distance  from the centre of the star with the strongest wind (assumed here to be star 1) to the shock is given by
\begin{equation}
  x_{\rm stag} = \frac{\sqrt{\eta}\,a}{1 + \sqrt{\eta}},
  \label{xstag}
\end{equation}
whilst the asymptotic opening angle $\theta_{\infty}$ of the shock cone (see Fig.\,\ref{fig:schema}) is the solution of the equation \citep{Canto}
\begin{equation}
  \theta_{\infty} - \tan{\theta_{\infty}} = \frac{\pi\,\eta}{\eta - 1}.
\end{equation}

In the most extreme case the dominant wind crashes directly onto the surface of the star with the less energetic wind. In the case of HD\,149404, this happens for values of $\eta$ exceeding 5.44, which correspond to $\theta_{\infty}$ values larger than about $120^{\circ}$. In our models, situations where $\eta$ exceeds this limiting value are treated as cases where observable X-rays are only emitted by the part of the shock region that is detached from the secondary surface. The emission zone along the wind interaction region was assumed to be restricted to cells located within a distance $r_{\rm max}$ from the centre of the star with the weaker wind (see Fig.\,\ref{fig:schema}).
\begin{figure}[h]
\begin{center}
\resizebox{8cm}{!}{\includegraphics{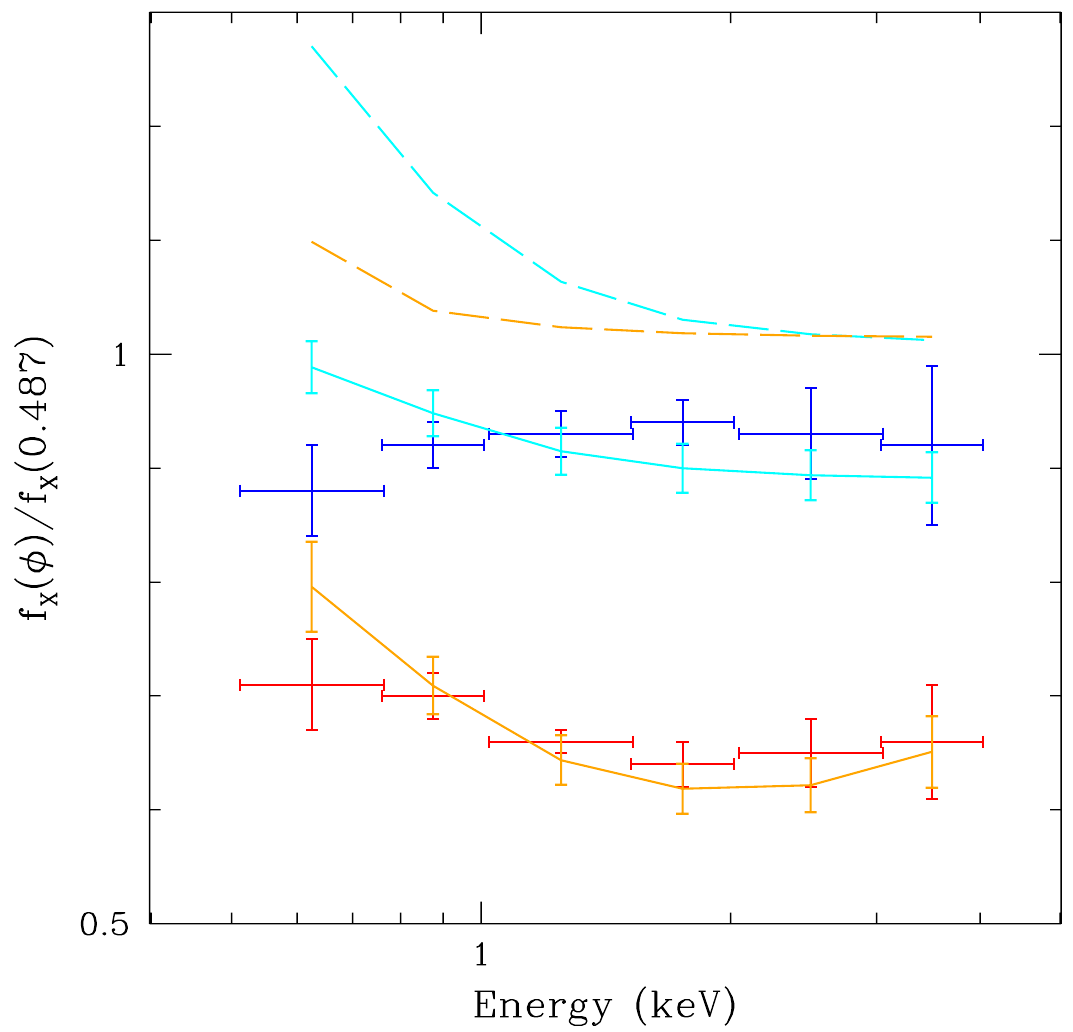}}
\end{center}  
\caption{Comparison between the observed flux ratios and the best-fit predicted values for different models tested in this work. The red and blue symbols with error bars correspond respectively to the measured flux ratios displayed in Fig.\,\ref{fig:fx}. The orange and cyan curves yield the best-fit theoretical models for the phases shown respectively by the red and blue symbols. The continuous lines assume the X-rays arise entirely from the wind interaction zone of HD\,149404 and using temperature and density profiles computed with the \citet{Canto} model for an adiabatic wind interaction zone (Sect.\,\ref{CWB}). The long-dashed lines yield the results for X-rays arising entirely from wind-embedded shocks intrinsic to the primary and secondary stars (Sect.\,\ref{EWS}). \label{fig:fxCanto}}
\end{figure}

For a given set of parameters, we can then use this model to predict the ratio of the X-ray fluxes that would be observed at phases 0.0 and 0.5 on the one hand, and phases 0.25 and 0.5 on the other hand. We finally compare these predictions to the observed values by evaluating
\begin{equation}
  \chi^2 = \sum_{\phi_j}\sum_{E_i} \frac{\left[\left(\frac{f_X(\phi_j,E_i)}{f_X(0.487,E_i)}\right)_{\rm obs} -\left(\frac{f_X(\phi_j,E_i)}{f_X(0.487,E_i)}\right)_{\rm mod}\right]^2}{\sigma^2\left(\frac{f_X(\phi_j,E_i)}{f_X(0.487,E_i)}\right)_{\rm obs}}
\end{equation}
for the two values of $\phi_j$ and the six values of $E_i$. 

We computed extensive grids of models in the five-dimensional parameter space ($i$, $\dot{M}_{\rm Prim}$, $v_{\rm Prim}$, $\dot{M}_{\rm Seco}$, $v_{\rm Seco}$). We tested various assumptions on the size of the emitting region, adopting $r_{\rm max}$ of $a/2$, $3\,a/4$, $a$, $3\,a$, $5\,a$, and no limitation on the size of the emitting zone. We further tested the assumptions that the wind momentum of the primary dominates over that of the secondary or that the reverse scenario holds.
\begin{table*}
  \caption{Stellar, wind, and orbital parameters adopted in the LIFELINE computations. \label{tab:lifeline}}
  \begin{center}
  \begin{tabular}{c c c c c c c}
    \hline
    & \multicolumn{2}{c}{Model I} & \multicolumn{2}{c}{Model II} & \multicolumn{2}{c}{Model III} \\
    \hline
    $i$ ($^{\circ}$) & \multicolumn{2}{c}{23} & \multicolumn{2}{c}{23} & \multicolumn{2}{c}{27} \\
    $a$ ($R_{\odot}$) & \multicolumn{2}{c}{78.6} & \multicolumn{2}{c}{78.6} & \multicolumn{2}{c}{67.6} \\
    \hline
    & Primary & Secondary & Primary & Secondary & Primary & Secondary \\
    \hline
    $M$ ($M_{\odot}$) & 42.2 & 25.5 & 42.2 & 25.5 & 26.9 & 16.2 \\ 
    $\dot{M}$ ($M_{\odot}$\,yr$^{-1}$) & $9.2\,10^{-7}$ & $3.3\,10^{-7}$ & $4.6\,10^{-7}$ & $1.7\,10^{-7}$ & $9.2\,10^{-7}$ & $3.3\,10^{-7}$ \\
    \hline
  \end{tabular}
  \end{center}
  \tablefoot{For each model the effective temperatures and radii are set to the values quoted in Table\,\ref{tab:paramCMFGEN}. $v_{\infty}$ is computed as $2.6\,v_{\rm esc}$, where $v_{\rm esc}$ is the escape velocity.}   
\end{table*}

Models where the secondary wind is assumed to dominate are significantly worse in reproducing the observed variations than models where the primary wind dominates. However, even for the latter models, the achieved quality of the best fits (i.e.\ those having the lowest $\chi^2$) is comparatively poor. The best results are obtained for $r_{\rm max} = a/2$ and values of $\eta$ close to 14 (see Fig.\,\ref{fig:fxCanto}) with $i = (23.0 \pm 0.9)^{\circ}$, $\dot{M}_{\rm Prim} = (1.9 \pm 0.4) \times 10^{-7}$\,$M_{\odot}$\,yr$^{-1}$, $\dot{M}_{\rm Seco} = (2.7 \pm 0.4) \times 10^{-7}$\,$M_{\odot}$\,yr$^{-1}$, $v_{\rm Prim} = (2539 \pm 184)$\,km\,s$^{-1}$, and $v_{\rm Seco} = 125$\,km\,s$^{-1}$. Figure\,\ref{fig:fxCanto} shows that even these best-fit models fail to reproduce correctly the energy dependence of the flux ratios, especially for the phase $\phi = 0.227$ observation. We also note that the best-fit parameters are quite different from the wind properties inferred by \citet{Raucq}. The low secondary wind velocity is not fully unexpected since in these models the primary wind crashes onto the secondary surface. What is more surprising is the fact that, in these models, both stars  have nearly identical mass-loss rates, with $\dot{M}_{\rm Seco}$ slightly exceeding $\dot{M}_{\rm Prim}$, whereas the opposite is expected (see Table\,\ref{tab:paramCMFGEN}). Whilst the secondary mass-loss rate would be close to the value expected from the models of \citet{Raucq}, the primary mass-loss rate would be five times below the expected value for a star of this spectral type. 

Owing to the small orbital separation of the stars, the plasma density in the wind interaction region is likely sufficient for radiative cooling to play a significant role \citep{SBP}. Under these circumstances, the shock-heated plasma cools before being advected out of the wind interaction region. This effect is amplified by the radiative inhibition of the wind velocity by the radiation pressure of the companion star \citep{SP94}. Radiative inhibition lowers the pre-shock velocities, thereby leading to cooler and denser post-shock plasma. In this case, the contribution of a given cell of gas to the X-ray luminosity is no longer proportional to the square of the plasma density, but instead scales linearly with the kinetic energy flux flowing into the shock \citep[e.g.][]{SBP,Kee}. Following \citet{Kee}, the emission integrated over both shock cooling layers is proportional to $\rho_0\,v_0^3$, where $\rho_0$ and $v_0$ are respectively the pre-shock wind density and the pre-shock wind velocity perpendicular to the shock. The power radiated in X-rays by a given cell is then evaluated as the product of $\rho_0\,v_0^3$ times the area of the cell along the shock surface. However, even this model provides only a first approximation to the effect of radiative cooling. As shown by \citet{Kee}, the occurrence of thin-shell instabilities \citep{Vishniac} not only leads to a significant reduction of the overall X-ray emission compared to the above scaling, but further yields a rather complex shape of the wind interaction zone. As a result, X-ray emission arises from a limited number of discrete patches associated with the tips and troughs of the unstable shear structure.

Another aspect that might be of relevance here is related to the impact of the orbital motion. In a short-period binary system such as HD\,149404, the Coriolis pseudo force could lead to a significant curvature and non-axial geometry of the wind interaction zone. Evidence of such a curvature was seen in the variability of the optical emission lines \citep{Rauw01,Naze}. It should be emphasised though that these optical emission lines probably arise from relatively distant parts of the wind interaction zone \citep{Rauw01,Naze}, whilst the X-ray emission is expected to form in the inner parts.

To determine how these two effects impact the putative X-ray emission from the wind interaction zone of HD\,149404, we used the LIFELINE code \citep{MosRau}, which combines the treatment of radiative driving and inhibition of \citet{SP94} with the formalism of \citet{ParPit} to treat the Coriolis deflection. LIFELINE uses the steady-state planar post-shock cooling layer model of \citet{Antokhin} to describe the density and temperature profiles of the plasma in the post-shock region that cools radiatively. The \citet{Antokhin} model is known to predict X-ray luminosities and plasma temperatures that are too high \citep[e.g.][]{FdP}. This is notably because of the above-described impact of thin-shell instabilities \citep{Vishniac,Kee}. Hence, the plasma temperatures and X-ray intensities predicted by LIFELINE should be seen as upper limits only. 

\begin{figure}[h]
\begin{center}
\resizebox{8cm}{!}{\includegraphics{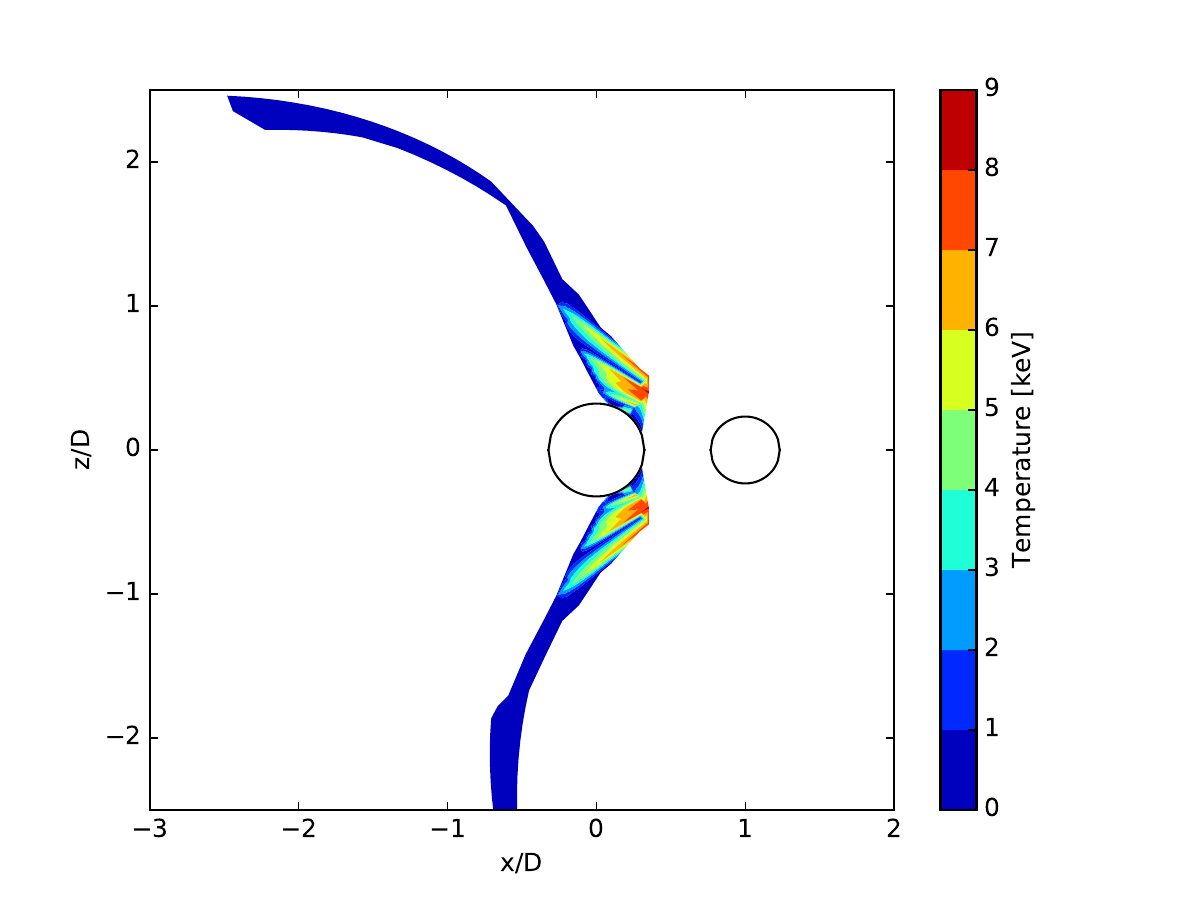}}
\end{center}
\begin{center}
\resizebox{8cm}{!}{\includegraphics{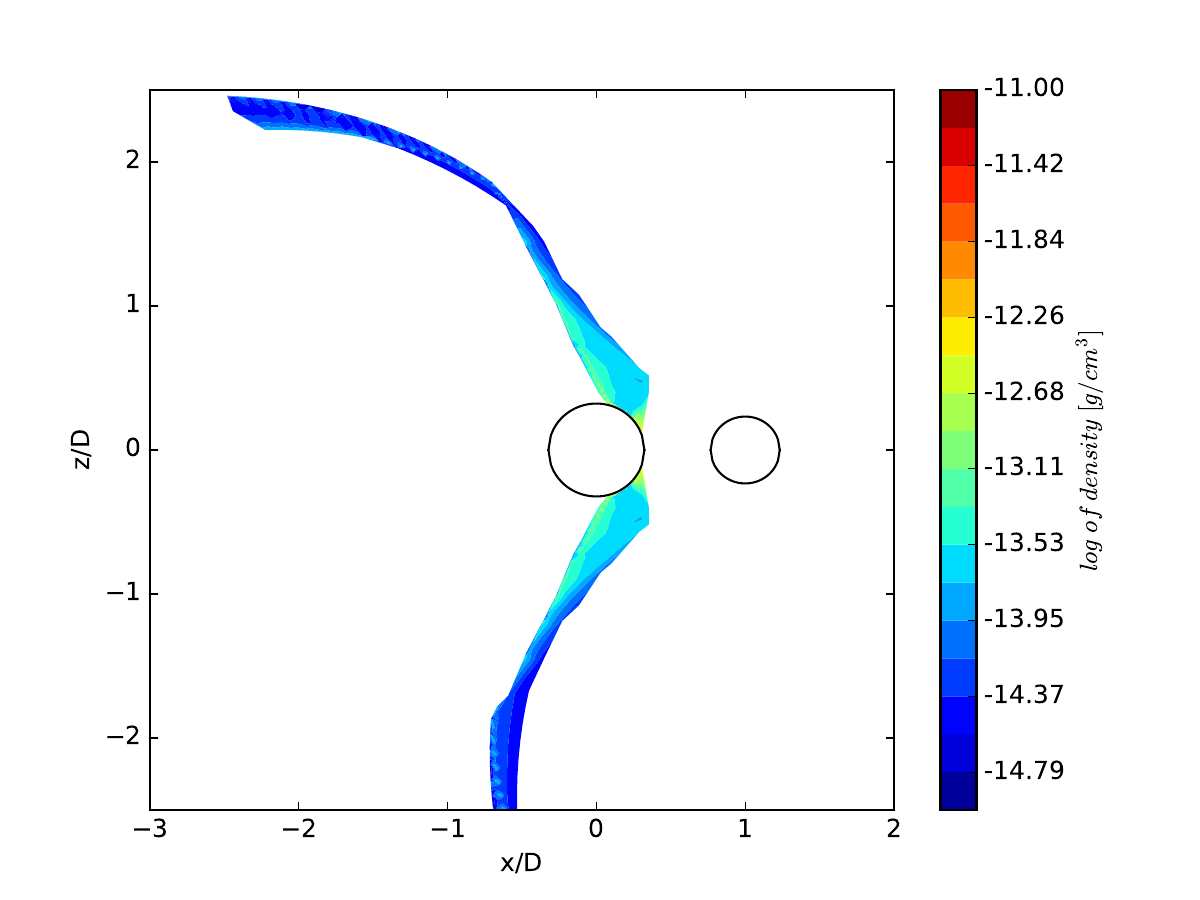}}
\end{center}
\caption{Theoretical temperature (top panel) and density (bottom panel) maps of the wind interaction zone in HD\,149404 computed with the LIFELINE code \citep{MosRau}, adopting the parameters of Model III (see Table\,\ref{tab:lifeline}). For this set of parameters, the wind of the primary star (on the right) crashes onto the surface of the secondary (on the left).\label{fig:temp}}
\end{figure}

We performed LIFELINE calculations for three sets of model parameters (see Table\,\ref{tab:lifeline}). Model I corresponds to the parameters of Table\,\ref{tab:paramCMFGEN}. Model II uses half mass-loss rates, whilst Model III considers a larger orbital inclination. In all three models, we observed a severe radiative inhibition effect. For instance, the primary wind reaches pre-shock velocities between 272\,km\,s$^{-1}$ (Model III) and 465\,km\,s$^{-1}$ (Model II). In models I and III, the primary wind crashes onto the secondary star despite the inclusion of radiative inhibition (see Fig.\,\ref{fig:temp}). The LIFELINE simulations predict that the hottest regions of the wind interaction zone are located in the zone where the shock detaches from the secondary surface. These regions are essentially unaffected by the Coriolis deflection (see Fig.\,\ref{fig:temp}). Therefore, the Coriolis deflection is unlikely to play a significant role in the interpretation of the variations in the X-ray emission of HD\,149404.

\subsection{An alternative description of the X-ray emission \label{EWS}}
Since HD~149404 does not appear to be a very bright and hard X-ray source, a significant part of its X-ray emission could arise from embedded wind shocks resulting from the intrinsic instability of the wind driving mechanism \citep{Lucy,OCR,Feld}. These shocks are expected to be distributed throughout the winds of the individual binary components. This mechanism is considered to produce the intrinsic X-ray emission of massive stars, the X-ray luminosity of which scales with the bolometric luminosity $L_{\rm bol}$ \citep[e.g.][and references therein]{Naze09,Handbook}. 

Unlike a single massive star, the stellar wind of a given component of a close binary system such as HD\,149404 cannot deploy equally in all directions. This is especially true in the case of a collision of the dominant wind with the photosphere of the companion. In such a situation the weaker wind only deploys over a solid angle limited by the wind interaction region. Using the notations of the \citet{Canto} model, the solid angle occupied by the weaker wind amounts to $2\,\pi\,(1 + \cos{\theta_{\infty}})$. For $\theta_{\infty} = 120^{\circ}$, which is the asymptotic shock angle predicted by the \citet{Canto} formalism for a wind-photosphere collision (see above) and which agrees well with the LIFELINE computations (see Fig.\,\ref{fig:temp}), the secondary's wind is able to deploy freely only over about one-quarter of the full $4\,\pi$ solid angle. Hence, compared to a single star of same $L_{\rm bol}$, the embedded wind shocks of the secondary star would contribute at most one-quarter of the X-ray emission expected from the canonical $\frac{L_X}{L_{\rm bol}}$ relation. Likewise, the primary wind can fully deploy only over a solid angle $2\,\pi\,(1 - \cos{\theta_{\infty}})$, that is three-quarters of the full solid angle. However, if the separation between the two stars is sufficient, that is to say, more than $a \simeq 2\,R_*$, embedded wind shocks might also develop in the part of the primary wind located between the two stars, ahead of the wind interaction zone. As we describe above, the wind velocity in that region is strongly impacted by radiative inhibition due to the secondary's radiation field \citep{SP94}, probably leading to weaker shocks and softer X-ray emission from the wind embedded shocks. To be conservative, we consider only the reduction of the intrinsic emission of the secondary star by a factor of four. This effect diminishes the intrinsic X-ray emission to 65\% of the value expected from the total bolometric luminosity of the binary system. Any effect on the primary's intrinsic emission would further reduce this value. In terms of $\log{f_X}$ this corresponds to a reduction in the expected flux due to wind embedded shocks of about $0.19$\,dex.

Observational studies of high-resolution X-ray spectra of single O-type stars showed that most of the X-ray emission arises from rather deep inside the stellar winds (typically within the inner $5\,R_*$), though there are clear differences in the emission formation regions according to the plasma temperature and the wind density \citep{Herve,lamCep}. The outer boundary of an emission region is ruled by the competition between the outward decreasing emission measure and the outward decreasing optical depth of the overlaying cool stellar wind. For instance, in the case of $\zeta$~Pup ($\log{\dot{M}} = -5.46$), \citet{Herve} derived emission regions that range between 1.5 and 38\,$R_*$ for the softest plasma component ($kT = 0.20$\,keV), 2.7 and 4.0\,$R_*$ for the $kT = 0.40$\,keV component and between 3.1 and 4.1\,$R_*$ for the hottest plasma component ($kT = 0.69$\,keV). In the case of $\lambda$~Cep ($\log{\dot{M}} = -5.85$), most of the X-ray emission was found to arise from regions between 1.1 and 2.5\,$R_*$ \citep{lamCep}. In both $\zeta$~Pup and $\lambda$~Cep, the volume filling factors of the hot gas were found to be near 0.01 \citep{Herve,lamCep}. The mass-loss rates of the stars in HD\,149404 are smaller than those of either $\zeta$~Pup or $\lambda$~Cep, resulting in lower density winds. Hence, we assume the intrinsic X-ray emission of those stars to form dominantly within about 2\,$R_*$ from the stellar centre.

To test the hypothesis that the entire observable X-ray emission arises from wind embedded shocks, we computed grids of models where three plasma components at temperatures found in the 3-T plasma models (see Table\,\ref{tab:fit3T}) are distributed across the wind of the primary and the part of the secondary wind contained within a shock cone attached to the secondary surface and with $\theta_{\infty} = 120^{\circ}$. Each plasma component was assumed to extend out to a radial distance of $r_{\rm out}$. The emission region was discretised into cells of 0.1\,$R_*$ in radial extension, and $2^{\circ}$ in both latitude and azimuth angles. For each of these emitting cells, we computed the emission measure and the optical depths across the wind using Eqs.\,(\ref{eqn1}) -- (\ref{eqn2}). We accounted for the fact that a given line of sight can intersect the shock cone up to two times, thus implying that for such lines of sight the winds of both stars contribute to the optical depth. We varied $r_{\rm out}$ in steps of 0.1\,$R_*$ between 1.05 and 2.45\,$R_*$, independently for each of the three plasma temperatures. We further considered volume filling factors of the two hotter components, which were varied between 1/16 and 8 times the volume filling factor of the coolest plasma component. Finally, we tested inclinations ranging from 23 to 31$^{\circ}$. We adopted the wind properties given in Table\,\ref{tab:paramCMFGEN}.

Although we explored a wide range of parameters, we found no combination of model parameters that successfully reproduced the observed variations (see the long-dashed lines in Fig.\,\ref{fig:fxCanto}). Instead, the results were much worse than those obtained assuming the X-ray emission to arise from the sole wind interaction zone. The models that came closest to the observations predict fluxes at all energies to be larger at $\phi = 0.0$ (secondary in front) than at $\phi = 0.5$ which is the reverse of what we observe. This situation stems from the combination of a slightly larger primary wind opacity and occultation effects of the secondary wind by the body of the secondary star at $\phi = 0.5$. Actually, the largest flux is predicted at quadrature phases (0.25 or 0.75), whilst observations  indicate this to be the case at $\phi = 0.5$. Moreover, the corresponding best-fit models predict a significant energy-dependence of the flux ratios, at odds with the observations. We also tested a grid of models where the primary mass-loss rate was reduced by a factor two, all other parameters remaining unchanged. These models did not improve the agreement with the observations. We thus conclude that the observed variations in the X-ray flux of HD\,149404 cannot be explained if the X-rays arise solely from embedded wind shock emission attenuated by an overlying smooth wind.   

\section{Discussion and conclusions\label{sect:conclusion}}
HD\,149404 displays variations in its X-ray emission at the $\sim 30$\% level. The system appears X-ray brightest when the O7.5\,I(f) primary is in front, and X-ray faintest when the ON9.7\,I secondary is in front. In a system with a circular orbit, such as HD\,149404, this behaviour is at first sight indicative of a phase-dependence of the optical depth of circumstellar column towards the X-ray source. This is reminiscent of the case of the O7 + Of$^+$/WN8ha 10.7-day period binary HDE\,228766 \citep{Rauw14}. For this latter system, the X-ray flux is severely absorbed at phases when the Of$^+$/WN8ha star is in front compared to phases when it is behind, which was consistently interpreted as the result of the higher opacity and higher density of the Of$^+$/WN8ha wind and the X-ray emission arising mostly from a wind-wind interaction zone. However, there are clear differences between the behaviours of HD\,149404 and HDE\,228766. First, the flux variations observed in HD\,149404 are of significantly lower amplitude than those of HDE\,228766. This could be due to the lower inclination of the orbit of HD\,149404, found to be in the range $23^{\circ}$ -- $31^{\circ}$ \citep{Rauw19}, whilst it was estimated in the range $54^{\circ}$ -- $61^{\circ}$ for HDE\,228766 \citep{Rauw14}, and to the lower mass-loss rate of the primary in HD\,149404 compared to the more extreme wind of the Of$^+$/WN8ha star. More surprisingly, the amplitude of the changes in the measured fluxes of HD\,149404 appears rather independent of photon energy (see Fig.\,\ref{fig:fx}), whereas a clear energy-dependence is expected for an absorption effect and was indeed observed for HDE\,228766 \citep{Rauw14}. The grey energy-dependence of the variations observed for HD\,149404 therefore suggests that the bulk of the variations instead stem from phase-dependent occultations of the X-ray emitting zone by the stellar bodies.

In our present study, we considered two possible physical origins for the X-ray emission: pure wind-wind collision and pure intrinsic emission arising from embedded wind shocks in the winds of the individual stars. Reproducing the observed variations as a function of energy is a challenge for both models. Because of their expected strong energy dependence, the models involving intrinsic emission perform significantly worse in reproducing the observed behaviour. Taken at face value, this suggests that the X-ray emission mostly arises from the innermost parts of the wind interaction   between the stars and with a location very close to the secondary star. However, with $\log{\frac{L_{\rm X}}{L_{\rm bol}}}$ ranging between $-7.13$ and $-6.97$, the overall level of the X-ray emission of HD\,149404 is comparatively low, unlike what one would expect in the presence of an X-ray bright wind-wind interaction.

The most plausible explanation for the system's grey variations of the X-ray emission could be that the stellar winds consist of partially optically thick fragments. \citet{Feld03} and \citet{Osk04} investigated a situation, based on the results of the hydrodynamic models of \citet{Feld}, where the stellar wind consists of randomly distributed fragments of compressed shells with a small lateral extension. The individual fragments could be optically thick, allowing photons to escape only via the lateral gaps between the fragments. Equation (35) in \citet{Osk04} provides an expression for the ensuing effective optical depth of such a porous wind. For optically thick fragments, this effective optical depth is indeed independent of photon energy, but depends on the average number of absorbing fragments along the radial direction $<N_r>$ and the radial extent of the wind region filled with such fragments. Wind porosity would allow soft X-ray photons arising inside the winds to escape, whereas such photons would be almost totally absorbed by a homogeneous wind. In the absence of a self-consistent formalism to predict the radial extent of the fragmented winds and the values of $<N_r>$, it is difficult to perform a quantitative comparison with the observations. However, since the highest X-ray emission level is observed around conjunction with the primary star in front, we can conclude that the primary wind should be more porous than the secondary wind, despite the higher mass-loss rate of the primary star \citep{Raucq}.
  
  In this context, it is also worth recalling that the BRITE photometry of HD\,149404 displayed a substantial red noise stochastic variability \citep{Rauw19}. This could reflect the presence of a sub-surface convection zone \citep{Can09} in at least one of the stars of the binary. Such a sub-surface convection zone could in turn trigger clumping, leading to a highly structured inner part of the wind.

  The most likely explanation of the phase behaviour of HD\,149404's X-ray emission thus appears to be that the stellar winds are clumpy and that these clumps are at least partially optically thick in the X-ray domain.   

\begin{acknowledgements}
The Li\`ege team acknowledges support from the Fonds National de la Recherche Scientifique (Belgium) notably under grant n$^o$ T.0192.19, and the Belgian Federal Science Policy Office (BELSPO) in the framework of the PRODEX Programme (contract HERMeS). We are greatful to Dr.\ Y.\ Naz\'e for constructive discussions. ADS and CDS were used for this research. 
\end{acknowledgements}

\appendix
\section{Optical depth along the line of sight \label{Appendix}}
To evaluate the optical depth from a given plasma cell towards the observer, we express the position of the plasma cell in the cylindrical $(p,z)$ coordinates of the star that is in front. For a wind velocity law
\begin{equation}
  v = v_{\infty}\,\left(1 - \frac{R_*}{r}\right),
\end{equation}
and defining
\begin{equation}
  \tau_{E,*} = \frac{\kappa_E(r=a/2)\,\dot{M}}{4\,\pi\,v_{\infty}\,R_*},
\end{equation}
the optical depth along the line of sight towards an observer located at $z \rightarrow \infty$ then becomes
\begin{equation}
  \tau_E(p,z) = \tau_{E,*}\,\ln{\left(\frac{z/R_*}{z/R_* - 1}\right)} \hspace*{5mm} \text{ for } p = 0,
  \label{eqn1}
\end{equation}
\begin{equation}
  \tau_E(p,z) = \frac{\tau_{E,*}}{\sqrt{1 - \frac{p^2}{R_*^2}}}\,\ln{\left(\frac{\alpha - x_2}{\alpha - x_1}\right)}  \hspace*{5mm} \text{ for } 0 < p < R_*,
\end{equation}
\begin{equation}
  \tau_E(p,z) = \frac{2\,\tau_{E,*}}{\alpha-1} \hspace*{5mm} \text{ for } p = R_*, \text{and}
\end{equation}
\begin{equation}
  \tau_E(p,z) = \frac{2\,\tau_{E,*}}{\sqrt{\frac{p^2}{R_*^2}-1}}\left[\frac{\pi}{2} - \arctan{\left(\frac{\frac{\alpha\,p}{R_*}-1}{\sqrt{\frac{p^2}{R_*^2}-1}}\right)}\right] \hspace*{3mm} \text{ for } p > R_*.
  \label{eqn2}
\end{equation}
In these equations
\begin{equation}
  \alpha = \tan{\left(\frac{\pi}{2}-\frac{1}{2}\,\arccos{\left(\frac{z}{\sqrt{p^2 + z^2}}\right)}\right)},
\end{equation}
\begin{equation}
  x_1 = \frac{R_*}{p} + \sqrt{\left(\frac{R_*}{p}\right)^2 - 1}, \text{ and }
\end{equation}
\begin{equation}
  x_2 = \frac{R_*}{p} - \sqrt{\left(\frac{R_*}{p}\right)^2 - 1}.
\end{equation}
For the X-ray emission arising from the wind interaction zone, only the cases $p \geq R_*$ are relevant, whereas for intrinsic emission arising from wind-embedded shocks, all situations described by Eqs.\,(\ref{eqn1}) -- (\ref{eqn2}) matter. 
\end{document}